\documentclass[11pt,a4paper]{article}
\pdfoutput=1
\usepackage{jheppub}
\usepackage{amsfonts,amssymb,epsfig,verbatim,mathbbol,amstext,amsmath,psfrag}
\usepackage{graphicx}
\usepackage[latin2]{inputenc}
\usepackage{t1enc}
\usepackage{amsmath}
\usepackage{subfigure}
\usepackage{dsfont}
\usepackage{slashed}
\usepackage{multirow}
\usepackage{lscape}
\usepackage{rotating}
\usepackage{wrapfig}
\usepackage{color}
\usepackage{array}

\usepackage{colordvi}
\usepackage{rotfloat}
\usepackage{amssymb}

\usepackage{cmap}

\renewcommand{\d}{\textmd{d}}
\newcommand{\be}{\begin{equation}}
\newcommand{\ee}{\end{equation}}

\newcommand{\Z}{\mathcal{Z}}

\newcommand{\F}{\mathcal{F}}
\newcommand{\E}{\mathcal{E}}
\renewcommand{\S}{\mathcal{S}}
\newcommand{\M}{\mathcal{M}}
\newcommand{\fracp}[2]{\frac{\partial #1}{\partial #2}}

\newcommand{\Regensburg}{Institute for Theoretical Physics, Universit\"at Regensburg, D-93040 Regensburg, Germany.}

\begin{document}

\title{QCD equation of state at nonzero magnetic fields in the Hadron Resonance Gas model}

\author[1]{G.~Endr\H{o}di,}
\affiliation[1]{\Regensburg}

\emailAdd{gergely.endrodi@physik.uni-regensburg.de}

\abstract{
The Hadron Resonance Gas (HRG) model is considered to study the QCD equation of state for the case of nonzero external magnetic fields.
Thermodynamic observables including the pressure, energy density, entropy density, magnetization and the speed of sound are presented as functions of the temperature and the magnetic field. 
The magnetization is determined to be positive, indicating that the hadronic phase of QCD is paramagnetic. The behavior of the speed of sound suggests that the deconfinement transition temperature is lowered as the magnetic field grows.
Moreover, a simple correspondence is derived, which relates the magnetic catalysis of the quark condensate to the positivity of the $\beta$-function in scalar QED.
}

\keywords{QCD equation of state, pressure, external magnetic field, background field method, magnetic catalysis}

\maketitle

\section{Introduction}

Quantum Chromodynamics (QCD) is the established theory of the strong interactions. QCD predicts that strongly interacting matter exhibits at least two very different forms: the hadronic phase at low temperatures $T$ and the quark-gluon plasma (QGP) phase at high $T$. Whereas the relevant degrees of freedom are color-singlet objects in the hadronic regime, they are colored particles in the quark-gluon plasma state. The transition between the QGP and the hadronic phase took place as the early universe expanded and cooled, about $10^{-5}$ seconds after the Big Bang. On the other hand, the same transition is also reproduced in heavy-ion collisions conducted in contemporary experiments at e.g. the Relativistic Heavy Ion Collider (RHIC) and the Large Hadron Collider (LHC). A particularly interesting aspect of the transition is how thermodynamic observables change as the system passes through the region separating the hadronic and QGP phase. The relations between these observables constitute the equation of state (EoS) of the system in equilibrium.

A striking observation made in heavy-ion collision experiments is that the flow of strongly interacting matter can be adequately described in terms of nearly-ideal relativistic hydrodynamic models, see, e.g. Refs.~\cite{Teaney:2000cw,Kolb:2003dz,Jacobs:2004qv,
Romatschke:2009im}. The dependence of the EoS on state parameters like the temperature $T$ and the chemical potential $\mu$ are necessary input to these models.
For a non-central heavy-ion collision, however, an external magnetic field $B$ is also generated by the spectators. Since the strength of this magnetic field reaches up to the hadronic scales~\cite{Skokov:2009qp,Bzdak:2011yy,Deng:2012pc}, it can have a significant impact on the properties of the transition and of the EoS. Similarly strong magnetic fields are expected to be present in dense neutron stars~\cite{Duncan:1992hi} and to be generated during the electroweak transition in the early universe~\cite{Vachaspati:1991nm}. In each of these cases, the interplay between the strong dynamics and the coupling to the external field can induce new and exciting phenomena. 
Examples include the the chiral magnetic effect~\cite{Kharzeev:2007jp,Fukushima:2008xe} or the decrease of the transition temperature with growing magnetic field~\cite{Bali:2011qj,Bali:2011uf}.

Thus, a clear theoretical understanding of the dependence of the EoS on $T$, $\mu$ and $B$ is desired. Most of our knowledge about the $B=\mu=0$ EoS comes from lattice Monte-Carlo studies, see e.g. Refs.~\cite{Bernard:2006nj,Cheng:2009zi,Borsanyi:2010cj}.
The inclusion of a finite chemical potential $\mu$ poses conceptual problems, when it comes to the lattice approach. On the contrary, nonzero magnetic fields are straightforwardly simulated using standard Monte-Carlo algorithms. Still, lattice results about the EoS for nonzero external magnetic fields are yet absent from the literature. 

On the other hand, there is another, remarkably simple approach, which can be used to access the low-temperature region of the EoS: the Hadron Resonance Gas (HRG) model. Within this model, the hadronic phase of QCD can be studied, even at nonzero chemical potentials or external magnetic fields. 
Comparison to lattice QCD results reveals that the HRG description gives a good approximation of thermodynamic observables even up to temperatures just below the transition region, both at zero chemical potential (see, e.g.,~\cite{Karsch:2003vd,Huovinen:2009yb,Borsanyi:2010cj}) and at nonzero chemical potential (see, e.g.,~\cite{Karsch:2003zq,Tawfik:2004sw,Allton:2005gk,Borsanyi:2012cr}), especially, if an exponential Hagedorn spectrum is also taken into account~\cite{Majumder:2010ik,NoronhaHostler:2012ug}. Still, the HRG description has not yet been employed for the case of nonzero magnetic fields.

In this paper, the hadronic EoS for nonzero magnetic fields is determined within the HRG model.
Sec.~\ref{sec:thermoB} is devoted to the discussion of thermodynamic relations in the presence of a magnetic field. In Sec.~\ref{sec:hrgdescr}, the HRG approximation is detailed, and the free energy for individual hadrons is calculated. The contributions from each hadron are summed up in Sec.~\ref{sec:results} to obtain the total free energy density, from which the whole equation of state is reconstructed. The $B$- and $T$-dependence of thermodynamic observables including the pressure, energy and entropy density, magnetization and speed of sound are calculated. Finally, in Sec.~\ref{sec:conc}, we conclude.

\section{Thermodynamics in an external magnetic field}
\label{sec:thermoB}

The quantity on the top of the hierarchy of thermodynamic relations is the thermodynamic potential -- which we refer to as free energy.
In terms of the partition function of the system, this free energy reads $\F=-T \log\Z$, and in the presence of a constant, external magnetic field $B$ is written as~\cite{kittel2004elementary,stanley1987introduction}
\be
\F = \E -T \S - B \M_B,
\label{eq:fundeq}
\ee
with $\E$ the energy\footnote{Note that $E$ in Eq.~(\ref{eq:fundeq}) denotes the total energy of the system, with the work necessary to maintain the constant external field also taken into account~\cite{kittel2004elementary}.}, $\S$ the entropy and $\M_B$ the magnetization. These observables satisfy the differential relations
\be
\fracp{\F}{T} = -\S,\quad\quad\quad\quad
\fracp{\F}{B} = -\M_B, \quad\quad\quad\quad
\fracp{\F}{V} = -p.
\label{eq:diffrels}
\ee
We also define here the corresponding densities as $\epsilon=\E/V$, $s=\S/V$, $f=\F/V$ and $m_B=\M_B/V$. In the thermodynamic limit, $V\to\infty$, differentiation with respect to the volume simplifies to a multiplication by $1/V$, and thus the pressure is given by
\be
p = -\frac{\F}{V} = -f = -(f^{\rm vac} + f^{\rm therm}),
\label{eq:pdef}
\ee
where we anticipated that the free energy separates into a vacuum and a thermal contribution, see Eq.~(\ref{eq:fvactherm}) below. 
From Eqs.~(\ref{eq:fundeq}) and~(\ref{eq:pdef}) the energy density is calculated as
\be
\epsilon=Ts+Bm_B -p.
\label{eq:edef}
\ee
Another observable of interest is the speed of sound $c_s$, which is defined in terms of differentials at constant $B$,
\be
c_s^2 = \left.\fracp{p}{\epsilon}\right|_{B} = \left.\fracp{p}{T}\right|_B \bigg/ \left.\fracp{\epsilon}{T}\right|_B.
\label{eq:Idef}
\ee
We remark that the definition~(\ref{eq:pdef}) leads to a pressure that is isotropic in space -- solely due to the assumption that the free energy $\F$ is an extensive quantity, and (in the thermodynamic limit) is proportional to $V$. As a result, the speed of sound will also be isotropic. Defining $p$ in terms of the diagonal elements of the stress energy tensor leads to anisotropic pressures (see, e.g., Ref.~\cite{Ferrer:2010wz}). 
A possible explanation for this apparent discrepancy was given in Ref~\cite{0022-3719-15-30-017}, in terms of a surface term to the stress energy tensor originating from the Lorentz force density. We remark that the isotropy properties of the pressure have also been discussed in Ref.~\cite{Bali:2013esa}, where compressions at fixed magnetic field and at fixed magnetic flux are distinguished.

\section{Free energy density in the HRG model}
\label{sec:hrgdescr}

In the HRG model~\cite{Dashen:1969ep}, the thermodynamic potential, Eq.~(\ref{eq:fundeq}), of the system is approximated in the thermodynamic limit, $V\rightarrow\infty$, by the partition function of a gas of non-interacting free hadrons and resonances~\cite{Venugopalan:1992hy}. 
Thus, the free energy density of the model is written as the sum of
independent contributions coming from non-interacting hadrons $h$,
\be
f = \sum_{h} d_h \cdot f_{h}\left(\{eB,T\},\{m_h,q_h/e,s_h,g_h\}\right),
\label{eq:hrglogz}
\ee
where each contribution $f_h$ depends on the external parameters (the magnetic field in elementary charge units $eB$ and the temperature $T$), and internal properties of the hadron (mass $m_h$, spin $s_h$, charge $q_h/e$ and gyromagnetic ratio $g_h$). Each hadron enters the sum with a certain multiplicity $d_h$. 
The hadrons taken into account extend from pions up to the $\Sigma^0$ baryon, as listed in the latest edition of the Particle Data Book~\cite{Beringer:1900zz}, and are tabulated in Table~\ref{tab:hadrons}. 
The experimental values for the gyromagnetic ratios in Ref.~\cite{Beringer:1900zz} are known only for a few hadrons, and only with large uncertainties (except for the proton and the neutron). Therefore we decided to take the gyromagnetic ratios to be $g_h=2 q_h/e$, as dictated by universal tree-level arguments~\cite{Ferrara:1992yc}. This corresponds to the assumption that the considered hadrons are point-like objects; e.g. neutral hadrons have $g_h=0$. A possible improvement of the method is to take into account the correct gyromagnetic ratios. The energy levels for $s=1/2$ and $s=1$ particles with anomalous magnetic moments have been discussed in Refs.~\cite{Goldman:1972vc,Tsai:1972iq}. These, however, lead to more complicated expressions for the free energies, for which the Landau sums cannot be performed in general.

\begin{table}[ht!]
\centering
\begin{tabular}{|l|c|c|c|c||l|c|c|c|c|}
\hline
hadron & $m (\textmd{GeV})$ & $|q/e|$ & $s$ & $d$ & hadron & $m (\textmd{GeV})$ & $|q/e|$ & $s$ & $d$ \\ 
\hline
\hline
$\pi^{\pm}$ & 0.135 & 1 & 0 & 2  & $p$ & 0.938 & 1 & 1/2 & 2  \\
$\pi^0$ & 0.135 & 0 & 0 & 1 & $n$ & 0.938 & 0 & 1/2 & 2  \\
$K^{\pm}$ & 0.495 & 1 & 0 & 2 & $\eta'$ & 0.958 & 0 & 0 & 1  \\
$K^0$ & 0.495 & 0 & 0 & 2 & $f_0$ & 0.980 & 0 & 0 & 1  \\
$\eta$ & 0.548 & 0 & 0 & 1 & $a_0$ & 0.980 & 0 & 1 & 1  \\ 
$\rho^{\pm}$ & 0.776 & 1 & 1 & 2 & $\phi$ & 1.020 & 0 & 1 & 1  \\
$\rho$ & 0.776 & 0 & 1 & 1 & $\Lambda$ & 1.116 & 0 & 1/2 & 1 \\  
$\omega$ & 0.782 & 0 & 1 & 1 & $h_1$ & 1.170 & 0 & 1 & 1 \\ 
$K_{*}^{\pm}$ & 0.892 & 1 & 1 & 2 & $\Sigma^\pm$ & 1.189 & 1 & 1/2 & 2 \\ 
$K_*^0$ & 0.892 & 0 & 1 & 2 &  $\Sigma^0$ & 1.189 & 0 & 1/2 & 1 \\ 
\hline
\end{tabular}

\caption{\label{tab:hadrons}List of hadrons and resonances taken into account in the HRG description.
}
\end{table}

\subsection{Energy levels}

To reconstruct the free energy in the low-temperature region using Eq.~(\ref{eq:hrglogz}), contributions from each particle type are summed up, with the assumption that the interaction between them is negligible. Let us therefore take a free relativistic particle with momentum $\mathbf{p}=(p_x,p_y,p_z)$ and mass $m$, in the presence of a magnetic field of magnitude $B$, pointing in the positive $z$ direction.
We consider the particle to have spin $s$ and charge absolute value $q$ (thus, in our notations $qB$ is always positive). The component of the spin in the direction of the magnetic field is a conserved quantity, and can assume the values $s_z=-s,-s+1,\ldots s$. The Landau levels are labeled by the index $k$. With these notations, the energy levels of a charged particle ($q>0$) in the presence of the magnetic field are given as~\cite{landau1977quantum},
\be
E(p_z,k,s_z) = \sqrt{p_z^2 + m^2 + 2qB \left(k+1/2 - s_z\right)},
\label{eq:energies}
\ee
while the energy levels for the neutral particle ($q=0$) are
\be
E_0(\mathbf{p})=\sqrt{\mathbf{p}^2 + m^2}.
\label{eq:energiesn}
\ee
As mentioned above, a generalization for the description of anomalous magnetic moments for $s=1/2$ and $s=1$ has been developed in Refs.~\cite{Goldman:1972vc,Tsai:1972iq}. 
In that approach, using a magnetic field-dependent gyromagnetic ratio, certain inconsistencies in the $s=1$ theory can be resolved. On the other hand, the description of the spin-$3/2$ theory in terms of Rarita-Schwinger fields~\cite{Rarita:1941mf} is known to exhibit non-causal behavior~\cite{Johnson:1960vt,Velo:1969bt}. Whether the simple formula~(\ref{eq:energies}) adequately describes the dispersion relation of a spin-$3/2$ particle is therefore not obvious. In fact, we will show in Sec.~\ref{sec:remarks} that -- unlike the other spin channels -- the $s=3/2$ sector would give a negative contribution to the pressure for any nonzero magnetic field, marking an instability in the theory.
For this reason we do not consider resonances with $s=3/2$ or higher in the model.

At arbitrary finite temperature, the free energy density (at vanishing chemical potentials) for a charged particle can be written as (see Ref.~\cite{kapusta2006finite} for the $B=0$ relation, and e.g. Ref.~\cite{Fraga:2008qn} for the analogous expression at $B\neq0$),
\be
f_{\rm c}(s) = \mp \sum_{s_z} \sum\limits_{k=0}^\infty \frac{qB}{2\pi} \int \frac{\d p_z }{2\pi} \left( \frac{E(p_z,k,s_z)}{2} + T\log(1 \pm e^{-E(p_z,k,s_z) / T}) \right),
\label{eq:chargedf}
\ee
where the lower sign corresponds to bosons ($s$ integer) and the upper one to fermions ($s$ half-integer)\footnote{Eq.~(\ref{eq:chargedf}) corresponds to a single charged particle. The contribution of the antiparticle is identical, and is taken into account in the total free energy density by considering the multiplicities of Table~\ref{tab:hadrons}.}.
The same for a neutral particle, on the other hand, is given by
\be
f_{\rm n}(s) = \mp \sum_{s_z} \int \frac{\d^3 \mathbf{p} }{(2\pi)^3} \left( \frac{E_0(\mathbf{p})}{2} + T\log(1 \pm e^{-E_0(\mathbf{p}) / T}) \right).
\label{eq:neutralf}
\ee

We separate the total free energy density into a vacuum ($T=0$) and a thermal part as
\be
f^{\textmd{vac}}(s)=\left.f(s)\right|_{T=0},\quad\quad\quad f^{\textmd{therm}}(s)=f(s)-f^{\textmd{vac}}(s).
\label{eq:fvactherm}
\ee
The vacuum terms in Eqs.~(\ref{eq:chargedf}) and~(\ref{eq:neutralf}) are ultraviolet divergent and need to be regularized. We use dimensional regularization with $d=1-\epsilon$. After separating the divergent contribution, the renormalization of the free energy density is carried out by subtracting the $B=0$ term, and by performing the renormalization of the pure magnetic energy $B^2/2$. As we will see, the latter is equivalent to renormalizing the elementary electric charge $e$~\cite{Schwinger:1951nm,Elmfors:1993bm,Dunne:2004nc,Andersen:2011ip}.
The vacuum free energy density has been obtained previously in e.g., Refs.~\cite{Schwinger:1951nm,Elmfors:1993bm,Chakrabarty:1996te,Ebert:1999ht,
Dunne:2004nc,Agasian:2008tb,Menezes:2008qt,Fraga:2008qn,Boomsma:2009yk,
Andersen:2011ip,Fraga:2012fs,Blaizot:2012sd}, however, the connection to electric charge renormalization has not been stated in all cases. 
In fact, it is a remarkable feature of the background field method, that electric charge renormalization -- and, accordingly, the coefficients of the $\beta$-function -- can be determined solely from how the free energy depends on an external magnetic field~\cite{Dunne:2004nc}.
In particular, this approach also gives insight into how the $\beta$-function can be related to the mass-dependence of the free energy -- and, thus, to the magnetic catalysis of the quark condensate (see Subsec.~\ref{sec:magncat}).
We remark that our renormalization prescription ensures, that the contribution to the free energy density diminishes as the mass of the particle increases, and thus the sum of Eq.~(\ref{eq:hrglogz}) in each spin sector will be convergent. 

On the other hand, the thermal part is explicitly finite (reflecting the fact that each divergence is independent of the temperature), and can be determined by numerical integration and summation. 

\subsection{Vanishing magnetic field}

In order to carry out the renormalization of the $B>0$ free energy density, first it is necessary to determine the $B=0$ contribution.
The vacuum free energy density at $B=0$ in $d=3-\epsilon$ dimensions is given by
\be
f^{\rm vac}(s,B=0) = \mp\frac{1}{2}(2s+1) \cdot \mu^\epsilon \int \frac{\d^{3-\epsilon} \mathbf{p}}{(2\pi)^{3-\epsilon}} \sqrt{\mathbf{p}^2 + m^2},
\ee
where the lower sign is for bosons, and the upper for fermions. Here the scale $\mu$ appeared to fix the dimension of the above expression to four.
To relate this integral to the nonzero $B$ case -- following Ref.~\cite{Menezes:2008qt} -- we rescale the momenta as $p\to p\sqrt{2qB}$, with $qB$ being an arbitrary dimensional scale. We denote $x=m^2/2qB$ and perform the integration using the formula~(\ref{eq:intformula}) to obtain
\be
f^{\rm vac}(s,B=0) = \mp\frac{1}{2} (2s+1) (2qB)^2 \frac{-1}{16\pi^2} \left(\frac{2qB}{4\pi \mu^2}\right)^{-\epsilon/2} \Gamma(-2+\epsilon/2) \, x^{2-\epsilon/2}.
\ee
We expand in $\epsilon$ using Eq.~(\ref{eq:expgamma}),
\be
f^{\rm vac}(s,B=0) =  \pm (2s+1) \frac{(qB)^2}{8\pi^2} x^2 \left[ \frac{1}{\epsilon} +\frac{3}{4} - \frac{\gamma}{2}-\frac{1}{2} \log\left(\frac{2qB}{4\pi \mu^2}\right) -\frac{1}{2}\log(x) \right].
\label{eq:vacB0}
\ee
Expressing this with $x=m^2/2qB$ cancels all $B$-dependence, of course.

\subsection{Nonzero magnetic field}
\label{sec:charged}

Let us now consider a particle with charge $q$ and spin $s$ in a magnetic field $B$. The energy levels are given by Eq.~(\ref{eq:energies}).
In terms of these levels, the vacuum free energy density is written using dimensional regularization as
\be
f^{\rm vac}(s) = \mp \frac{1}{2}\sum_{k=0}^{\infty} \sum_{s_z} \frac{qB}{2\pi} \cdot\mu^\epsilon \int\frac{\d^{1-\epsilon}p_z}{(2\pi)^{1-\epsilon}} \sqrt{p_z^2+m^2+2qB(k+1/2-s_z)},
\ee
where again the upper sign corresponds to fermions and the lower to bosons, and $\mu$ is the scale related to dimensional regularization.
We abbreviate $a=1/2-s_z$ and integrate in $p_z$ (using formula~(\ref{eq:intformula})) to get
\be
f^{\rm vac}(s) = \pm \frac{(qB)^2}{8\pi^2} \left(\frac{2qB}{4\pi \mu^2}\right)^{-\epsilon/2} \Gamma(-1+\epsilon/2) 
\sum_a  \zeta(-1+\epsilon/2,x+a),
\ee
where the sum over $k$ was converted into a Hurwitz $\zeta$ function, see Eq.~(\ref{eq:Hurwitzdef}). Now expanding in powers of $\epsilon$, using Eq.~(\ref{eq:expgamma}), we obtain
\be
f^{\rm vac}(s) = \pm \frac{(qB)^2}{8\pi^2} \sum_a \left[
\left(-\frac{2}{\epsilon}+\gamma +\log\left(\frac{2qB}{4\pi\mu^2}\right)-1\right) \left(-\frac{1}{12} -\frac{(x+a)^2}{2}+\frac{x+a}{2}\right) - \zeta'(-1,x+a)
\right].
\ee
To calculate the change in the free energy density due to the magnetic field, we subtract the $B=0$ contribution, Eq.~(\ref{eq:vacB0}),
\be
\Delta f^{\rm vac}(s) = \pm \frac{(qB)^2}{8\pi^2} \sum_a \left[
\left(\frac{2}{\epsilon}-\gamma-\log\left(\frac{2qB}{4\pi\mu^2}\right)+1\right) \left( \frac{1}{12}-\frac{a}{2}+\frac{a^2}{2}\right)  - \zeta'(-1,x+a) - \frac{x^2}{4} + \frac{x^2}{2}\log(x)
\right],
\label{eq:funren}
\ee
where we used that $\sum_a (a-1/2)=0$.

This expression is still divergent, as it contains the purely magnetic field-dependent term $\sim B^2/\epsilon$. In order to cancel this divergence, we have to redefine the free energy density by including in it the energy density of the magnetic field, $B^2/2$~\cite{Schwinger:1951nm,Elmfors:1993bm,Dunne:2004nc,Andersen:2011ip}. The divergence is then absorbed into the renormalization of the electric charge, and simultaneously, into that of $B$,
\be
\Delta f^{\rm vac,r} = \Delta f^{\rm vac} + \frac{B^2}{2},
\quad\quad\quad B^2 = Z_q B_r^2,
\quad\quad\quad q^2 = Z_q^{-1} q_r^2,
\quad\quad\quad q_rB_r=qB,
\label{eq:renormcharged}
\ee
with the renormalization constant
\be
Z_q=1\mp\frac{q_r^2}{8\pi^2} \sum_a\left( \frac{2}{\epsilon}-\gamma -\log\left(\frac{m_\star^2}{4\pi\mu^2}\right) \right)\left( \frac{1}{6}-a+a^2\right),
\label{eq:renormconst}
\ee
which, for a spin-$1/2$ particle,
\be
Z_q^{\rm spinor}=1+\frac{1}{2}\beta_1^{\rm spinor} \,q_r^2 \left( -\frac{2}{\epsilon}+\gamma +\log\left(\frac{m_\star^2}{4\pi\mu^2}\right) \right),
\quad\quad\quad\quad \beta_1^{\rm spinor}=\frac{1}{12\pi^2},
\label{eq:renormconstQED}
\ee
reproduces the well-known expression\footnote{The factor $1/2$ in front of $\beta_1^{\rm spinor}$ is canceled, if the contribution of the antiparticle is also taken into account. To convert to cutoff regularization, see Eq.~(\ref{eq:schemerelation}).} (see, e.g., Ref.~\cite{itzykson2006quantum}), with the leading coefficient of the spinor QED $\beta$-function appearing in front of the divergence. Similarly, for a spin-$0$ particle we get
\be
Z_q^{\rm scalar}=1+\frac{1}{2}\beta_1^{\rm scalar} \,q_r^2 \left( -\frac{2}{\epsilon}+\gamma +\log\left(\frac{m_\star^2}{4\pi\mu^2}\right) \right),
\quad\quad\quad\quad \beta_1^{\rm scalar}=\frac{1}{48\pi^2},
\label{eq:renormconstscalarQED}
\ee
where the scalar QED $\beta$-function coefficient enters (cf. Ref.~\cite{Dunne:2004nc}). Here $m_\star=m$ is a constant, which is fixed to the physical mass of the particle (see discussion in Subsec.~\ref{sec:magncat}). The renormalization of Eq.~(\ref{eq:renormcharged}) then leads to
\be
\Delta f^{\rm vac,r}(s) = \frac{B_r^2}{2} \mp \frac{(qB)^2}{8\pi^2} \sum_a \bigg[
\zeta'(-1,x+a) 
- \frac{x^2}{2}\log(x)+\frac{x^2}{4} -
\left(\frac{1}{12}-\frac{a}{2}+\frac{a^2}{2}\right)\left(\log(x)+1\right)
\bigg],
\label{eq:fch}
\ee
which can be rewritten in a compact form in the original variables,
\be
\begin{split}
\Delta f^{\rm vac,r}(s) = \frac{B_r^2}{2} \mp \frac{(qB)^2}{8\pi^2} \bigg[
&\sum_{s_z} \zeta'\left(-1,x+1/2-s_z\right) \\
&+(2s+1)\cdot \left( \frac{x^2}{4}-\frac{x^2}{2}\log(x)+\frac{\log(x)+1}{24} \left(1 - 4 s(s+1)\right) \right) 
\bigg].
\end{split}
\label{eq:pgeneral}
\ee
We remark that instead of using dimensional regularization, the free energy density in the $s=0$ and $s=1/2$ sectors can also be calculated using Schwinger's proper time formalism~\cite{Schwinger:1951nm}, which gives the same results~\cite{Ebert:1999ht,Dunne:2004nc}. The renormalized free energies of Eq.~(\ref{eq:pgeneral}) are given for $s=0$, $1/2$ and $1$ in App.~\ref{app2}.

Note that the renormalization constant $Z_q$ is such, that the expansion of the renormalized $\Delta f^{\rm vac,r}$ in the magnetic field at $\mathcal{O}(B^2)$ exactly equals $B_r^2/2$ -- in accordance with the expectation that the free energy, at $T=0$ and at small magnetic fields, comes exclusively from the external field itself.
Indeed, using the asymptotic behavior, Eq.~(\ref{eq:zeta_asympt}), of the Hurwitz $\zeta$ function, it is easy to check that the only contribution to $\Delta f^{\rm vac,r}$ at $\mathcal{O}(B^2)$ is the pure magnetic term, and the small $B$ (large $x$) limit of the square parentheses in the second contribution of Eq.~(\ref{eq:pgeneral}) is zero. Note that this also ensures, that the large mass limit of the matter contribution vanishes. 
Different renormalization schemes have also been used in the literature. For example, the scheme employed in Ref.~\cite{Menezes:2008qt} corresponds to dropping all mass-independent terms in Eq.~(\ref{eq:funren}), whereas in the one used in Ref.~\cite{Fraga:2012fs}, an additional $\sim(qB)^2\zeta'(-1,a)$ finite term is also subtracted. 
These schemes are connected to that of Eq.~(\ref{eq:renormconst}) by a finite renormalization, which would produce a renormalized vacuum free energy of the form $B_r^2/2+\mathcal{O}(B^2)$ -- instead of defining the total quadratic term to be $B_r^2/2$.
Accordingly, the different renormalization in Refs.~\cite{Menezes:2008qt,Fraga:2012fs} leads to a $\Delta f^{\rm vac,r}$ (and thus, also to a magnetization $m_B$) that grows logarithmically with the mass -- instead of approaching zero for $m\to\infty$, as expected on physical grounds. The prescription~(\ref{eq:renormconst}) is the only choice, for which the total quadratic term is $B_r^2/2$, and the large mass limit of the free energy (and the magnetization) vanishes.

If more particles (possibly with different masses, charges and spins) are present in the system, the renormalization constant $Z_q$ is extended to absorb the divergences coming from the interaction of each particle with $B$. The pure magnetic energy $B_r^2/2$ will be unchanged, independently of the number and properties of these particles. 
Note that the remaining part of the free energy, $\Delta f^{\rm vac,r}-B_r^2/2$, is induced by the interaction of the magnetic field with virtual hadrons present in the quantum vacuum. This is indeed not a classical effect, since it cancels in the entropy (which is written as the temperature-derivative of the free energy, Eq.~(\ref{eq:diffrels})), but is of purely quantum mechanical origin. In particular, $\Delta f^{\rm vac,r}$ can be represented as an infinite sum of loop diagrams with even number of external photon lines (with special momenta, such that these photons correspond to the external magnetic field). In this representation, the leading $\mathcal{O}((qB)^4)$ term, for example, is given by the scattering of two photons through a virtual charged hadron loop.

\subsection{Renormalization and magnetic catalysis}
\label{sec:magncat}

Let us consider the mass-dependence of the vacuum free energy density.
In this respect, one has to carefully distinguish between the actual mass $m$ of the hadron, and the fixed mass $m_\star$ that appears in the renormalization prescription, Eq.~(\ref{eq:renormconst}). Writing out the variables explicitly, the first relation of Eq.~(\ref{eq:renormcharged}) reads
\be
\Delta f^{\rm vac,r}(m,m_\star) = \Delta f^{\rm vac}(m) + \frac{B_r^2}{2} \,Z_q(m_\star).
\label{eq:separation}
\ee
The renormalization ensures that the total renormalized free energy density up to $\mathcal{O}(B^2)$ equals the pure magnetic contribution $B_r^2/2$ for a hadron of mass $m=m_\star$,
\be
\left.\Delta f^{\rm vac,r}(m,m_\star)\right|_{m=m_\star} = \frac{B_r^2}{2} + \mathcal{O}(B^4).
\label{eq:afterren}
\ee
The renormalization prescription (i.e., $m_\star$) does not change, if the mass of the hadron is varied -- in this sense the employed scheme is mass-independent.

Let us now consider the up quark condensate at $T=0$, which is given in terms of the vacuum free energy density as
\be
\bar uu \equiv -\frac{\partial f^{\rm vac,r}}{\partial m_u},
\ee
where $m_u$ is the mass of the up quark.
We can write the change in the condensate due to the magnetic field through the hadron sigma terms ($\partial (m^2)/\partial m_u$) as
\be
\Delta\bar uu = -\frac{\partial \Delta f^{\rm vac,r}(m,m_\star)}{\partial (m^2)} \cdot \frac{\partial (m^2)}{\partial m_u}=
-\frac{\partial \Delta f^{\rm vac}(m)}{\partial (m^2)} \cdot \frac{\partial (m^2)}{\partial m_u}
\label{eq:pbpdef}
\ee
where we inserted Eq.~(\ref{eq:separation}).
One can easily check using Eq.~(\ref{eq:funren}) for $\Delta f^{\rm vac}$, that the result is a positive condensate $\Delta \bar uu>0$ of $\mathcal{O}(B^2)$, in agreement with the well-known magnetic catalysis mechanism (see, e.g., Refs.~\cite{Gusynin:1995nb,Bali:2012zg}).\footnote{A very similar argument applies to a different observable as well. 
Recently it was shown that the QCD magnetization separates into spin- and orbital angular momentum-related contributions, and the spin term $\bar u\sigma_{\mu\nu} u$ has been determined on the lattice~\cite{Bali:2012jv}. In fact, $\bar u\sigma_{\mu\nu} u$ can be written as the derivative of $\Delta f^{\rm vac,r}$ for a spin-$1/2$ quark, with respect to the gyromagnetic ratio $g$ (replace $a=1/2-s_z$ with $a=1/2-g/2\cdot s_z$ in the calculation of Subsec.~\ref{sec:charged}). The renormalization prescription of Eq.~(\ref{eq:renormconst}), on the other hand, contains the fixed $g_\star=2$ and, thus, does not contribute to the derivative with respect to $g$. Again, the result is of $\mathcal{O}(B^2)$, in agreement with Ref.~\cite{Bali:2012jv} -- in contrast to the total vacuum free energy density that we calculate here, which is always of the form $B_r^2/2+\mathcal{O}(B^4)$.}

There is a further consequence of the separation, Eq.~(\ref{eq:separation}).
Let us consider the case of charged pions, with $m=m_\pi$, for which the sigma term is calculated using the Gell-Mann-Oakes-Renner relation,
\be
m^2 f_\pi^2 = (m_u+m_d)\,\bar u u_{B=0} \quad\quad\to\quad\quad
\frac{\partial (m^2)}{m_u} = \frac{\bar uu_{B=0}}{f_\pi^2},
\label{eq:gmor}
\ee
where $\bar u u_{B=0}$ is the zero-field up quark condensate and $f_\pi$ the chiral limit of the pion decay constant. 
Inserting Eq.~(\ref{eq:afterren}) into Eq.~(\ref{eq:separation}), we get,
\be
\frac{B_r^2}{2} +\mathcal{O}(B^4) = \Delta f^{\rm vac}(m) + \frac{B_r^2}{2} \,Z_q^{\rm scalar}(m).
\ee
Now we differentiate this equation with respect to $m_u$. Since the left hand side is independent of the mass up to $\mathcal{O}(B^2)$, we can express the condensate, using the right hand side of Eq.~(\ref{eq:pbpdef}), as
\be
\Delta\bar uu = \frac{B_r^2}{2}\, \frac{\partial Z_q^{\rm scalar}(m)}{\partial (m^2)} \cdot \frac{\partial (m^2)}{\partial m_u} + \mathcal{O}(B^4)= \frac{1}{2}\,\beta_1^{\rm scalar}\, (qB)^2\,  \frac{\bar u u_{B=0}}{m^2 f_\pi^2} + \mathcal{O}(B^4),
\label{eq:newrelation}
\ee
where we inserted the expression~(\ref{eq:renormconstscalarQED}) of the scalar QED renormalization constant for a pair of positively and negatively charged pions, and employed Eq.~(\ref{eq:gmor}). Using the value of $\beta_1^{\rm scalar}=1/(48\pi^2)$, the result reproduces the chiral perturbation theory formula~\cite{Agasian:2001hv,Cohen:2007bt} for the charged pion contribution to the condensate, up to $\mathcal{O}(B^2)$.
Altogether, this shows that the magnetic catalysis of the condensate is a direct consequence of the actual form of the electric charge renormalization constant in scalar QED. In particular, the fact that scalar QED is not asymptotically free ($\beta_1^{\rm scalar}>0$), implies that the quark condensate undergoes magnetic catalysis at $T=0$, and increases (to leading order) quadratically with growing $B$.

In the following, we will not consider the condensate, but only the free energy itself, for which it is not necessary to distinguish between $m$ and $m_\star$. Moreover, we will exclude the pure magnetic term $B_r^2/2$ from consideration, since it gives no information about the response of hadrons to the external field. Therefore, the vacuum pressure will be of $\mathcal{O}(B^4)$. The superscript $r$ denoting the renormalized free energy will also be dropped.

\subsection{Stability and spin channels}
\label{sec:remarks}

We notice that the formula~(\ref{eq:pgeneral}) is only well-defined for values of $x=m^2/2qB$ for which
\be
x+\frac{1}{2} - s > 0 
\quad\quad \to \quad\quad
(2s-1)\, qB < m^2.
\ee
For any spin $s\ge1$, this constraint gives a critical magnetic field $B_c$, where the theory breaks down. Clearly, as $B_c$ is approached, the assumptions of the model -- in particular, that hadrons are point-like particles -- become incorrect. This implies that one is restricted to $x>1/2$ for the $s=1$ and $x>1$ for the $s=3/2$ channel. The critical magnetic fields corresponding to the $\rho^{\pm}$ hadron is $eB_c(\rho^{\pm}) = m_\rho^2\approx 0.6\textmd{ GeV}^2$. This observation forms the basis of the idea of a superconducting vacuum at high magnetic fields $B>B_c$~\cite{Chernodub:2010qx}.

\begin{figure}[ht!]
\centering
\includegraphics*[width=8.5cm]{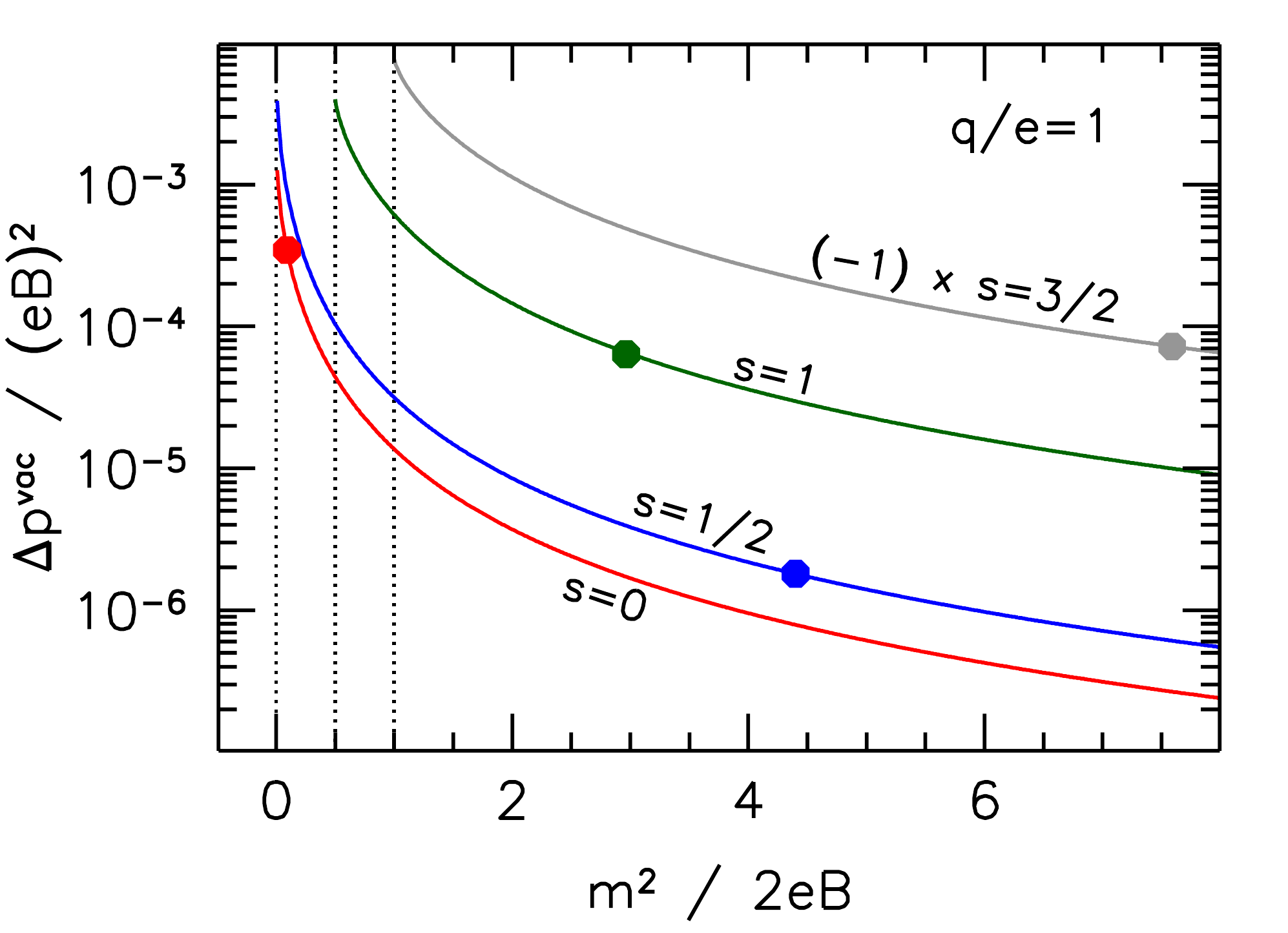}
\vspace*{-0.15cm}
\caption{Magnetic field dependence of the vacuum pressure in various spin channels for charge $q/e=1$. The dots represent the corresponding values for a $\pi^{\pm}$ (red), a proton (blue), a $\rho^{\pm}$ (green) and a $\Delta^{\pm}$ particle (gray) at a magnetic field $eB=0.1 \textmd{ GeV}^2$. The dashed lines represent lower bounds on $m^2/2eB$ for each spin sector. The contribution from the $\Delta^{\pm}$ has been multiplied by $-1$.}
\label{fig:xdep}
\end{figure}

In Fig.~\ref{fig:xdep} we plot the dependence~(\ref{eq:pgeneral}) of the vacuum pressure $\Delta p^{\rm vac}=-\Delta f^{\rm vac}$ on $m^2/2qB$ in the four spin channels $s=0\ldots3/2$, with $q/e=1$. We find that the vacuum pressure for the $s=3/2$ channel is negative (and has been multiplied by $-1$ in the figure), whereas the other three are positive.
In fact, above $s=1$, the sign of $\Delta p^{\rm vac}$ starts to alternate, with spin-integer hadrons contributing positively and spin-half-integer hadrons negatively to the pressure. The $s=1/2$ channel seems to be the only exception to this rule. 
The colored dots in Fig.~\ref{fig:xdep} represent the corresponding lowest-lying particles at $eB=0.1\textmd{ GeV}^2$. Although for this magnetic field, the charged pion is seen to dominate, as the magnetic field grows, at some point the $\Delta^{\pm}$ would clearly take over, turning the total vacuum free energy density $f^{\rm vac}$ positive, and, accordingly, the $T=0$ pressure negative. This indicates an instability, which suggests that the HRG model in terms of the dispersion relation~(\ref{eq:energies}) is not applicable for $s=3/2$.

\section{Results}
\label{sec:results}

\begin{figure}[ht!]
\centering
\includegraphics*[width=8.5cm]{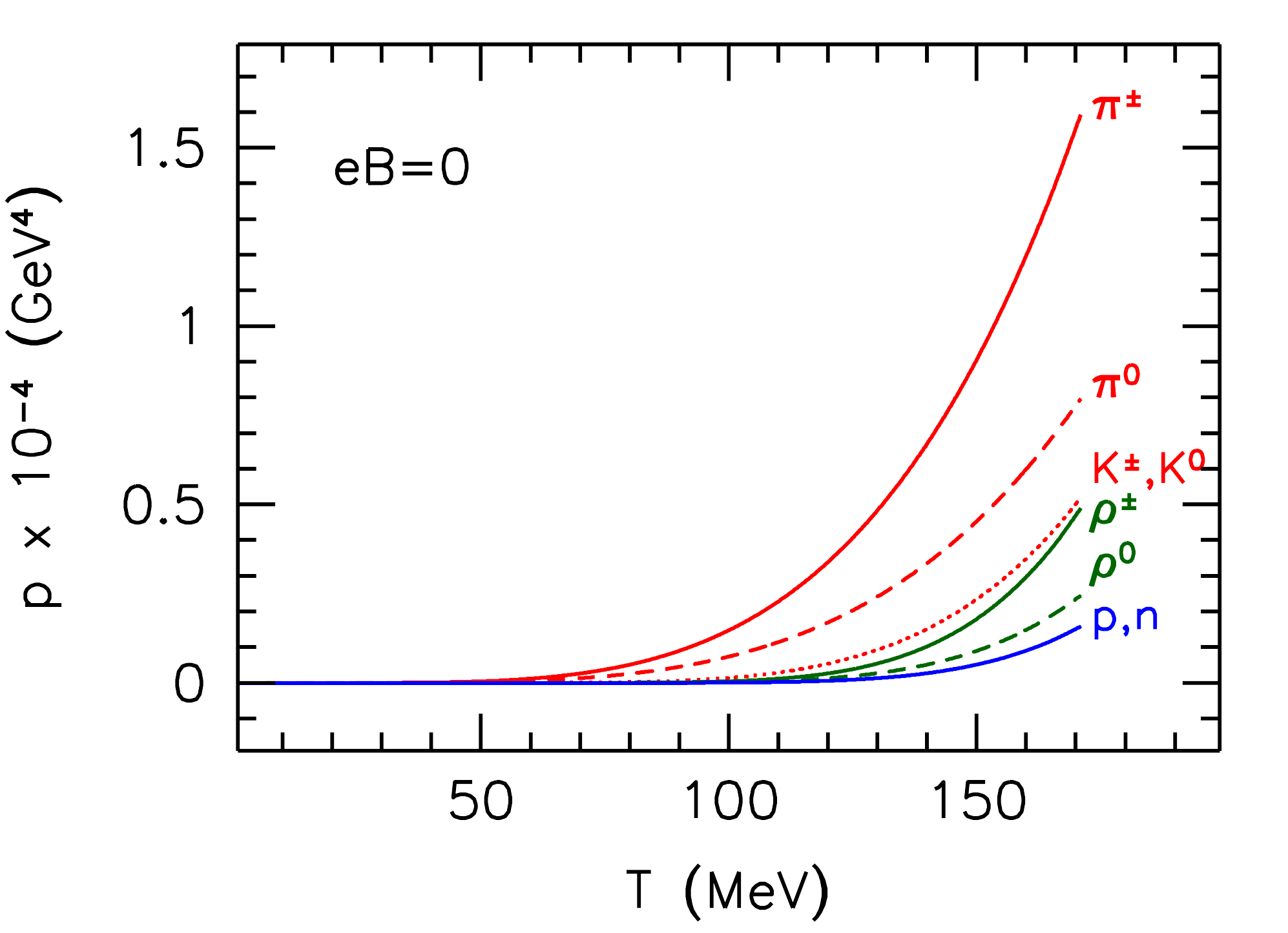}
\includegraphics*[width=8.5cm]{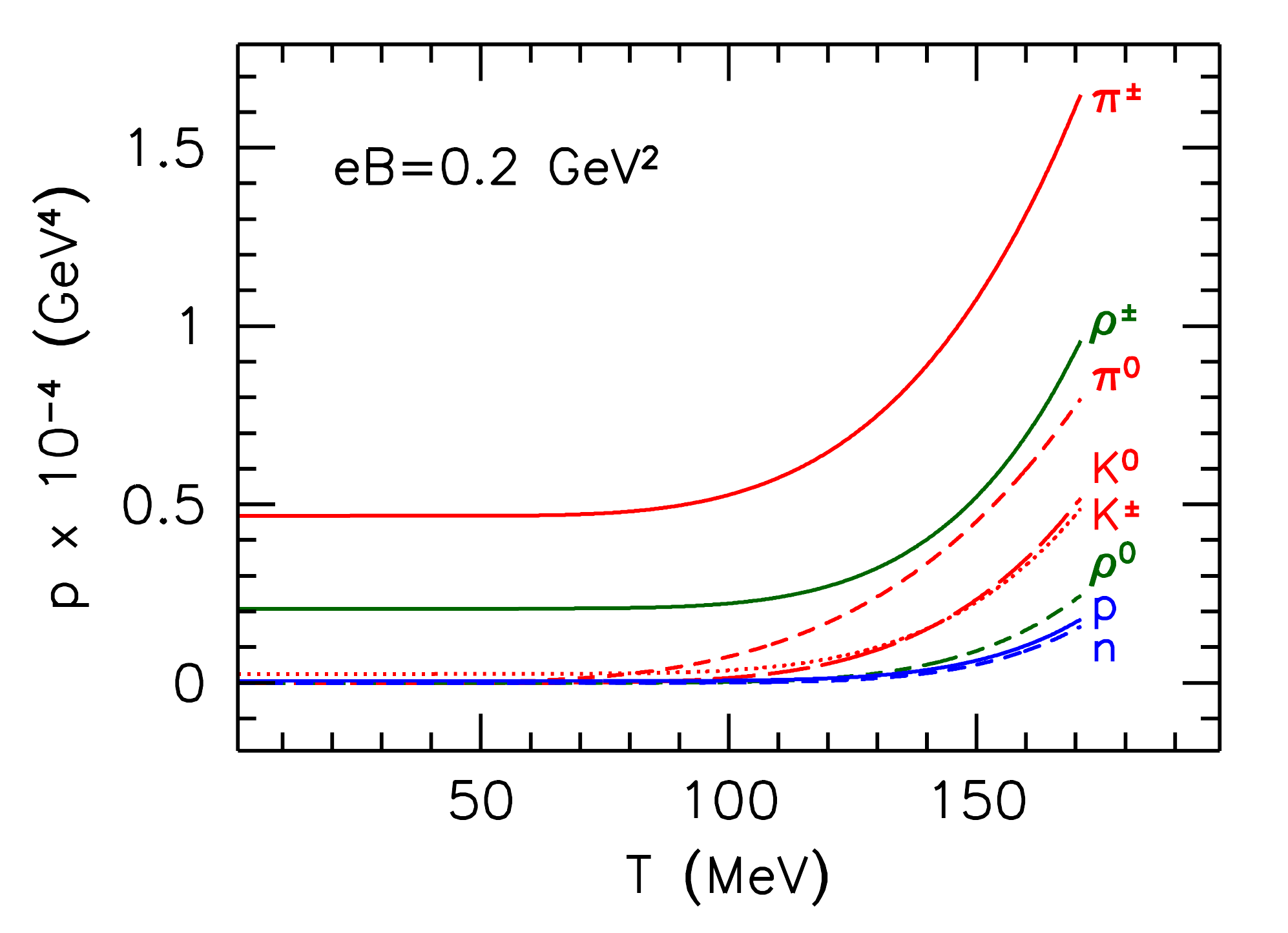}
\vspace*{-0.15cm}
\caption{Individual contributions to the HRG pressure as a function of the temperature for $eB=0$ (left panel) and $eB=0.2 \textmd{ GeV}^2$ (right panel). Note the change in the contribution of charged particles, especially $\pi^{\pm}$ and $\rho^{\pm}$, between the two panels.}
\label{fig:contributions}
\end{figure}

We will make use of the thermodynamic relations~(\ref{eq:diffrels})-(\ref{eq:Idef}) to determine the equation of state for nonzero magnetic fields.
At $B=0$, it is customary to normalize the free energy density, or, the pressure, by $T^4$. As the magnetic field is switched on, the $T=0$ pressure is in general nonzero, and thus $p/T^4$ diverges as $T$ decreases. 
We remark that the pressure at $T=0$ is nonzero even after dropping the pure magnetic energy $B_r^2/2$ in Eq.~(\ref{eq:pgeneral}), as $(qB)^4$ contributions are still present.
Therefore, in the following we will plot the pressure and other EoS-related observables in physical units, without a normalization by $T^4$. In Fig.~\ref{fig:contributions}, the individual contributions of various hadrons to the pressure are shown as functions of the temperature, for $B=0$ (left panel) and for $eB=0.2 \textmd{ GeV}^2$ (right panel). In the low temperature region ($T<100 \textmd{ MeV}$), the pressure is dominated by pions for vanishing magnetic field. As $B$ increases, this dominance is lost, as vacuum contributions in the other spin channels -- most importantly, for the $\rho^{\pm}$ hadron -- arise.

Since, roughly speaking, the effective mass of charged particles is $m^2_{\rm eff}=m^2+qB(1-2s)$ (see Eq.~(\ref{eq:energies})), it increases with $B$ for $s=0$ hadrons, but decreases for $s=1$ particles. Accordingly, the thermal part of the pressure -- which contains the Boltzmann weights $\exp(-m_{\rm eff}/T)$ -- is larger for $\rho^{\pm}$, whereas it smaller for $\pi^{\pm}$, as compared to the $B=0$ case. This effect is also visible in the temperature dependence of the $\pi^{\pm}$ and $\rho^{\pm}$ contributions in the right panel of Fig.~\ref{fig:contributions}.
We note moreover, that in our approach neutral particles are not affected by the magnetic field, since their gyromagnetic ratios are set to zero.

\begin{figure}[ht!]
\centering
\vspace*{-.2cm}
\includegraphics*[width=8.5cm]{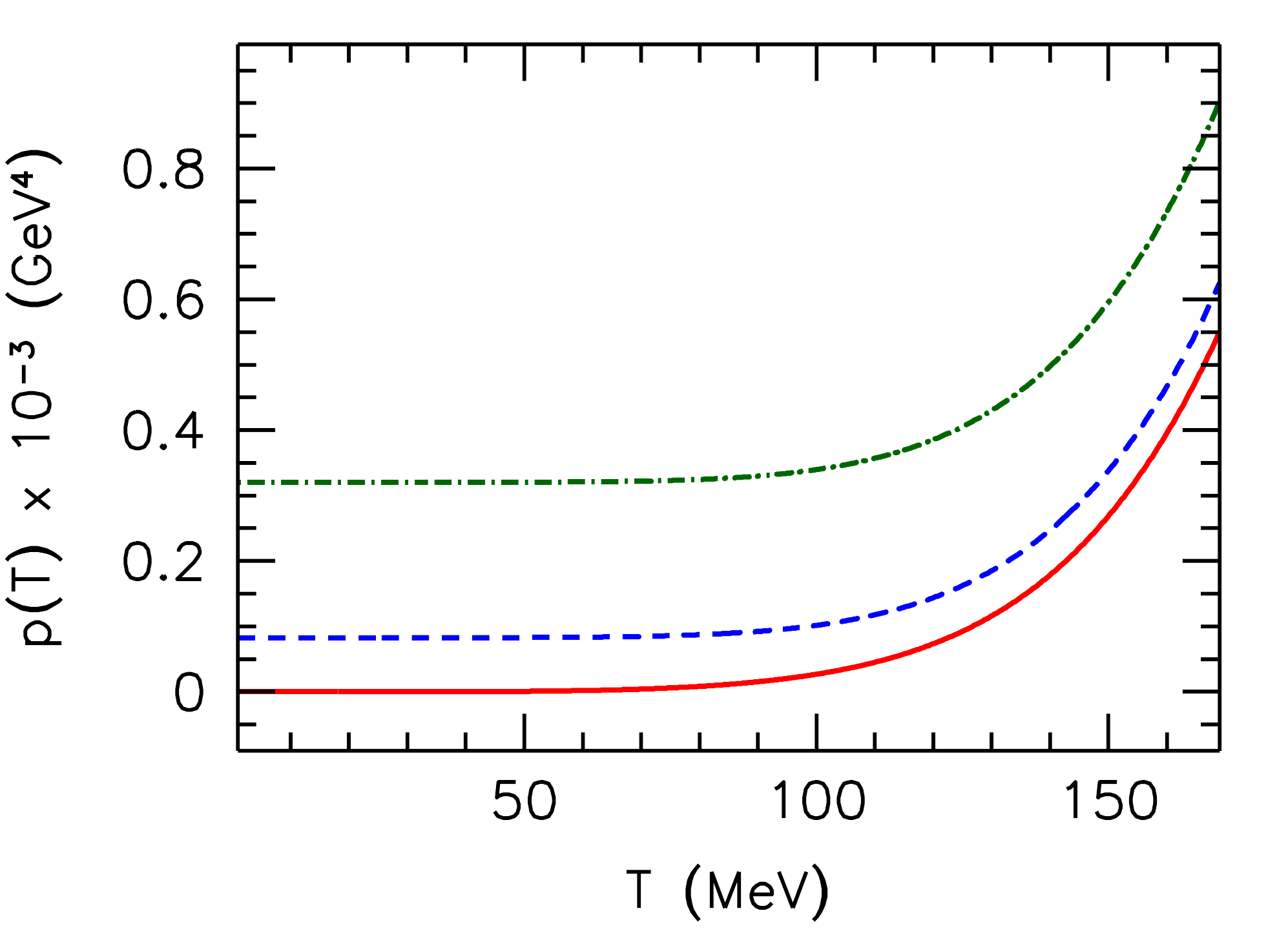}
\hspace*{.4cm}
\includegraphics*[width=8.5cm]{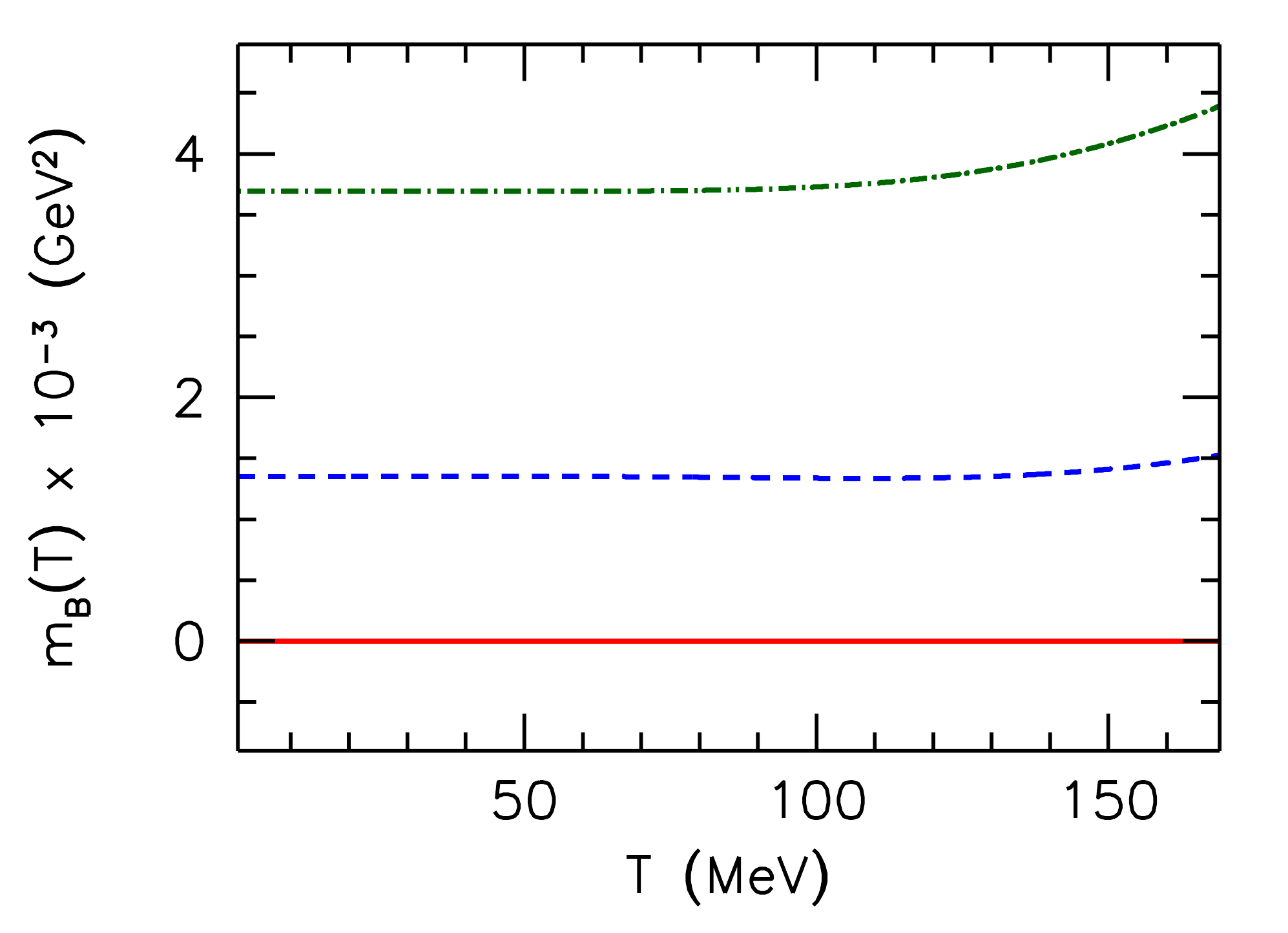}
\vspace*{-.3cm}
\includegraphics*[width=8.5cm]{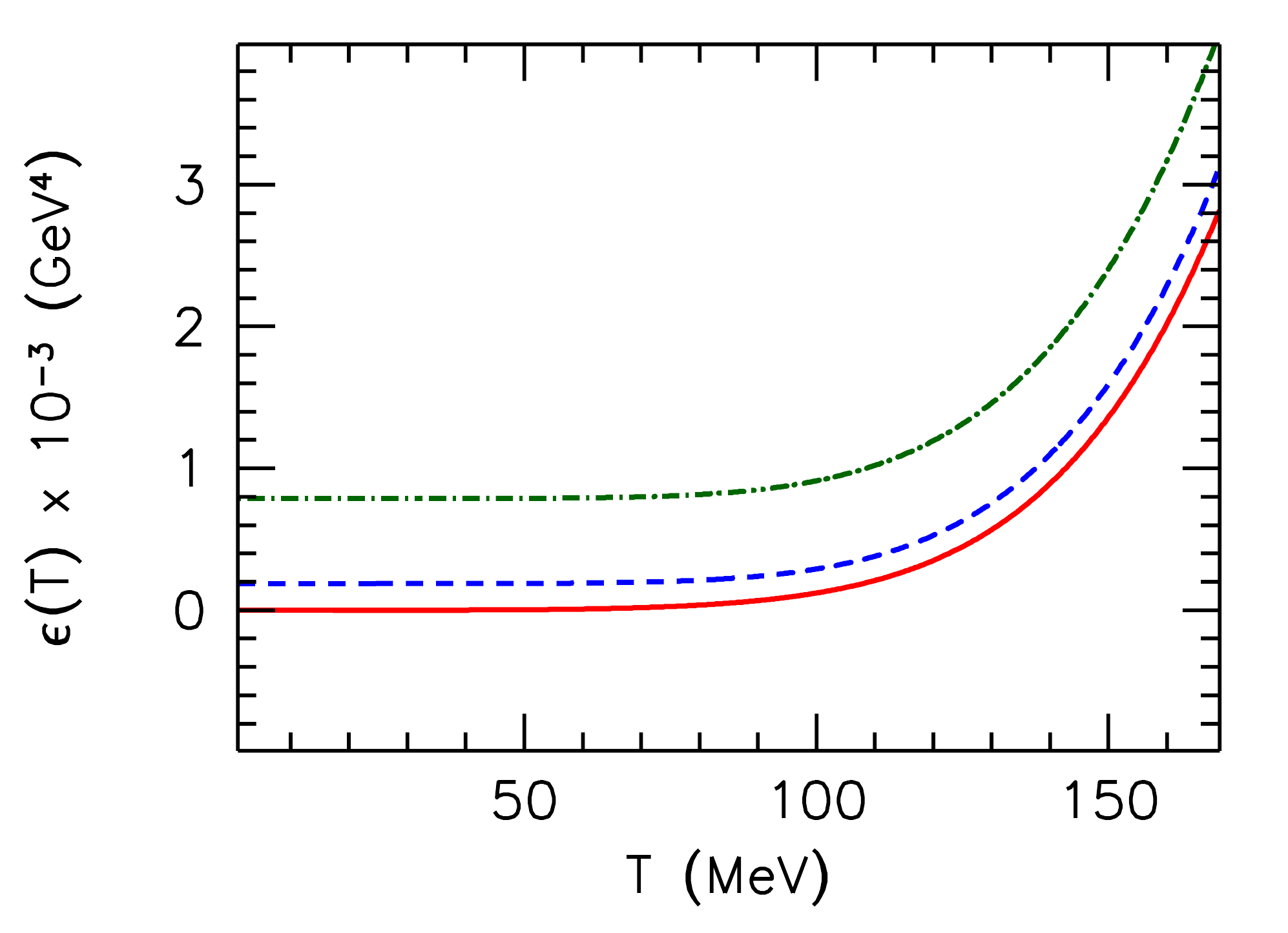}
\hspace*{.4cm}
\includegraphics*[width=8.5cm]{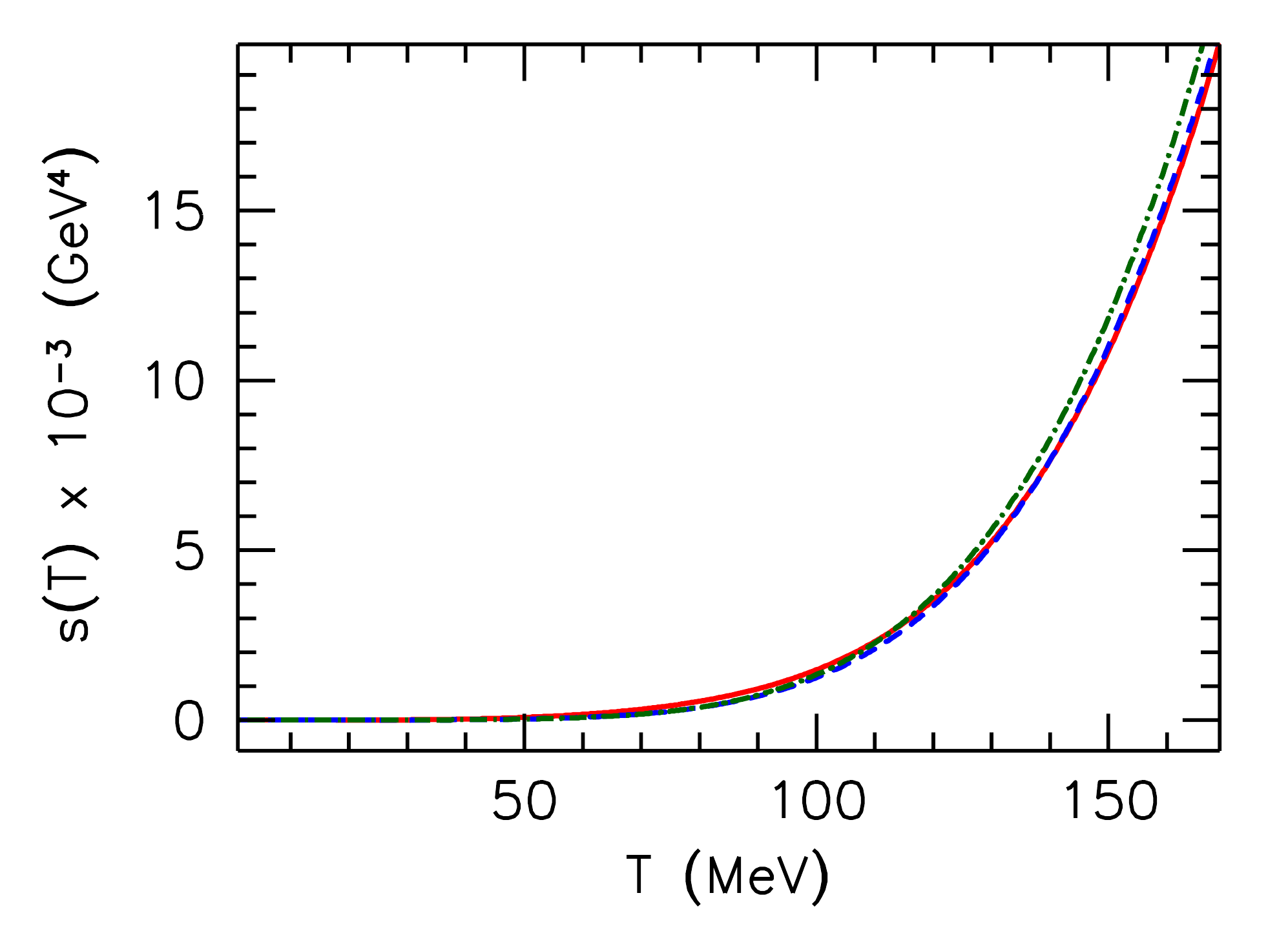}
\includegraphics*[width=8.5cm]{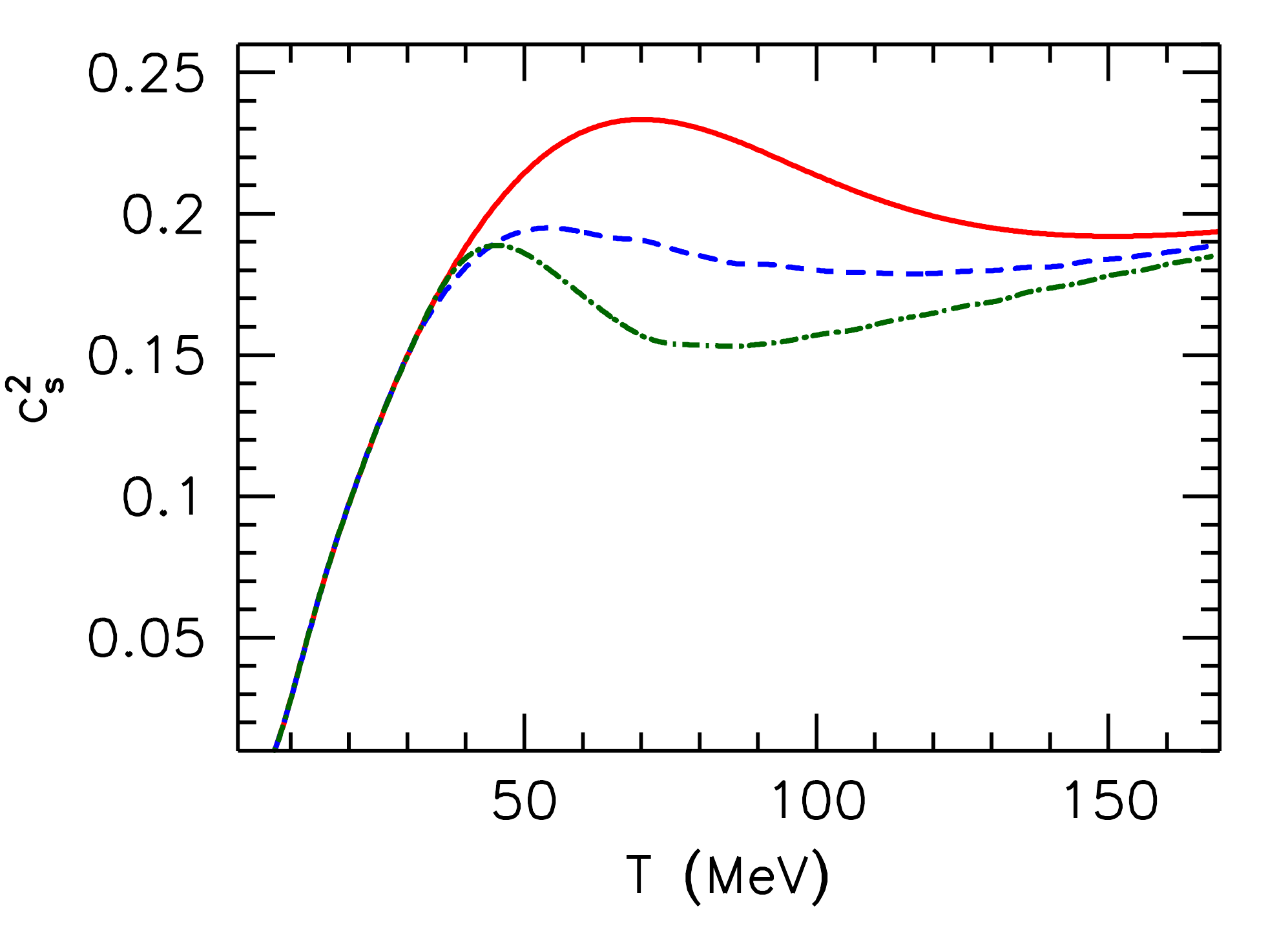}
\vspace*{-0.3cm}
\caption{The equation of state in the HRG model. Shown are (from left to right and downwards) the pressure, the magnetization, the energy density, the entropy density and the speed of sound squared as functions of the temperature, for $eB=0$ (solid red lines), $eB=0.2\textmd{ GeV}^2$ (dashed blue) and $eB=0.3\textmd{ GeV}^2$ (dot-dashed green).}
\label{fig:EoS}
\end{figure}

We proceed by performing the sum over hadrons to determine the dependence of thermodynamic observables on $B$ and $T$. These observables are the pressure $p$, the energy density $\epsilon$, the entropy density $s$, the magnetization $m_B$ and the speed of sound $c_s$ (for their definition, see Sec.~\ref{sec:thermoB}). In Fig.~\ref{fig:EoS}, these quantities are plotted as functions of the temperature for $eB=0$, $0.2\textmd{ GeV}^2$ and $0.3\textmd{ GeV}^2$. 
The pressure is increased by the magnetic field in the hadronic sector, and, correspondingly, the magnetization is positive, indicating a paramagnetic hadronic phase -- with interesting implications regarding the deconfinement transition temperature, based on large $N_c$ arguments~\cite{Fraga:2012ev}.

Since the entropy is written as the derivative of $p$ with respect to $T$, it is insensitive to the vacuum contribution, and vanishes at zero temperature. Moreover, even at nonzero temperatures, $s$ barely changes with $B$, in the range of magnetic fields under consideration. 
A much more pronounced signal can be seen in the speed of sound, which exhibits a dip, moving towards lower temperatures as $B$ is increased. The minimum position of $c_s^2$ is one possible definition of the deconfinement transition temperature $T_c$ (see, e.g. Ref.~\cite{Borsanyi:2010cj}). The behavior of the speed of sound thus indicates that $T_c$ decreases as the external magnetic field grows, in agreement with the recent lattice results~\cite{Bali:2011qj,Bali:2011uf}.

\section{Conclusions}
\label{sec:conc}

In this paper, we have developed a Hadron Resonance Gas model to study the QCD equation of state for nonzero magnetic fields. 
Using the renormalization properties of the free energy density, we have derived the relation Eq.~(\ref{eq:newrelation}), which connects the scalar QED $\beta$-function and the mass-dependence of the quark condensate. This correspondence explains the well-known magnetic catalysis phenomenon, in terms of the renormalization group running of the scalar QED coupling, thereby relating two, seemingly very different concepts.

We proceeded by investigating the individual contributions from hadrons to the pressure, and observed, that pions no longer dominate the low-temperature region if the magnetic field exceeds $B\gtrsim 0.2 \textmd{ GeV}^2$, thus, it is essential to take into account higher-lying resonances -- especially the $\rho^{\pm}$ hadron. By summing up the total pressure and considering its $B$-dependence, the magnetization is determined to be positive, showing that the QCD vacuum at low temperatures is paramagnetic. Using the magnetization and the pressure, the whole equation of state is reconstructed. The behavior of the speed of sound suggests, that the deconfinement transition temperature is lowered as the magnetic field is increased. 
We stress that the results obtained in the HRG model are reliable only at low temperatures and low magnetic fields. The limitation in the temperature is obviously given by the transition to the quark-gluon plasma phase, and as lattice results show, the HRG approximation at $B=0$ is reliable up to $T\approx 130-150\textmd{ MeV}$. Concerning the magnetic field, $eB < m_\rho^2 \approx 0.6\textmd{ GeV}^2$ must be fulfilled for the description to be consistent. The magnetic field also has to be small enough such that the assumption of neglecting $s=3/2$ hadrons (see Subsec.~\ref{sec:remarks}) is valid. 
The lattice determination of the EoS at nonzero magnetic fields is necessary to test the limitations of the HRG approach, and to confirm our findings.

\acknowledgments
I am grateful for enlightening discussions with Jens Andersen, Gerald Dunne and Igor Shovkovy about electric charge renormalization. I would also like to thank Gunnar Bali, Falk Bruckmann, Eduardo Fraga, S\'andor Katz, Andreas Sch\"afer and K\'alm\'an Szab\'o for essential remarks and a careful reading of the manuscript, and Jorge Noronha and Let\'icia Palhares for useful discussions. I thank the ECT* for their hospitality during the workshop `QCD in strong magnetic fields', where this idea was born.
This work was supported by the EU (ITN STRONGnet 238353). 

\appendix

\section{Formulae}

In this appendix we summarize the formulae that were used in the derivation of the vacuum free energy density in Subsec.~\ref{sec:charged}.

\begin{itemize}
\item
For integration in $d$ dimensions we use (see, e.g., Ref.~\cite{peskin1995introduction})
\be
\int_{-\infty}^{\infty} \frac{\d^d p}{(2\pi)^d} \sqrt{p^2 + M^2} = \frac{1}{(4\pi)^{d/2}} \frac{\Gamma(-1/2-d/2)}{\Gamma(-1/2)} (M^2)^{1/2+d/2},
\label{eq:intformula}
\ee
and $\Gamma(-1/2)=-2\sqrt{\pi}$.

\item
The expansion of the $\Gamma$ function around some negative integers is given by
\be
\Gamma(-1+\epsilon/2) = -\frac{2}{\epsilon} + \gamma - 1 +\mathcal{O}(\epsilon), \quad\quad\quad
\Gamma(-2+\epsilon/2) = \frac{1}{\epsilon} - \frac{\gamma}{2} + \frac{3}{4} +\mathcal{O}(\epsilon),
\label{eq:expgamma}
\ee
where $\gamma$ is the Euler constant.

\item
The Hurwitz $\zeta$ function is defined as
\be
\sum_{k=0}^{\infty} \frac{1}{(x+k)^z} = \zeta(z,x),
\label{eq:Hurwitzdef}
\ee
with the expansion~\cite{dlmf}
\be
\zeta(-1+\epsilon/2,x) \approx -\frac{1}{12} -\frac{x^2}{2} + \frac{x}{2} + \frac{\epsilon}{2} \zeta'(-1,x) + \mathcal{O}(\epsilon^2),
\ee
and asymptotic behavior~\cite{dlmf},
\be
\zeta'(-1,x) = \frac{1}{12} -\frac{x^2}{4} + \left( \frac{1}{12} -\frac{x}{2} + \frac{x^2}{2}\right) \log(x) +
\mathcal{O}(x^{-2}).
\label{eq:zeta_asympt}
\ee

\item
Writing the same integral in dimensional regularization with parameter $\epsilon$ and scale $\mu$, and in cutoff regularization with cutoff $\Lambda$,
\be
\begin{split}
-(\mu^2)^{\epsilon/2} \int_0^{\infty} \frac{\d s^{1+\epsilon/2}}{s} e^{-m^2s} &= -\frac{2}{\epsilon}+\gamma+\log\left(\frac{m^2}{\mu^2}\right) + \mathcal{O}(\epsilon),\\
-\int_{1/\Lambda^2}^{\infty} \frac{\d s}{s} e^{-m^2s} &= \gamma+\log\left(\frac{m^2}{\Lambda^2}\right) + \mathcal{O}(1/\Lambda^2),
\end{split}
\ee
we note that the two schemes are related as
\be
-\frac{2}{\epsilon} +\log\left(\frac{m^2}{\mu^2}\right) \leftrightarrow \log\left(\frac{m^2}{\Lambda^2}\right).
\label{eq:schemerelation}
\ee

\item
The Hurwitz $\zeta$ function at shifted second argument fulfills
\be
\zeta'(-1,x+1)=\zeta(-1,x)+x \log(x),
\label{eq:zetaxp1}
\ee
which holds since the derivatives of the two sides are equal, and $\zeta'(-1,0)=\zeta'(-1,1)$.

\end{itemize}

\section{Renormalized free energies}
\label{app2}

Here we give the renormalized free energy, Eq.~(\ref{eq:pgeneral}), for spins $s=0$, $1/2$ and $1$, without the pure magnetic energy $B_r^2/2$,
\be
\begin{split}
\Delta f^{\rm vac,r}(0) &= \frac{1}{8\pi^2} (qB)^2\left[
\zeta'\left(-1,x+1/2\right) 
+\frac{x^2}{4}-\frac{x^2}{2}\log(x)+\frac{\log(x)+1}{24}
\right],\\
\Delta f^{\rm vac,r}(1/2) &= \frac{-1}{4\pi^2}(qB)^2 \left[
\zeta'\left(-1,x\right)+\frac{x}{2}\log(x)  
+\frac{x^2}{4}-\frac{x^2}{2}\log(x)-\frac{\log(x)+1}{12}
\right],\\
\Delta f^{\rm vac,r}(1) &= \frac{3}{8\pi^2} (qB)^2\,\bigg[
\zeta'\left(-1,x-1/2\right)+\frac{1}{3}(x+1/2)\log(x+1/2) + \frac{2}{3}(x-1/2)\log(x-1/2) \\
&\hspace*{2.3cm}+\frac{x^2}{4}-\frac{x^2}{2}\log(x)-7\,\frac{\log(x)+1}{24}
\bigg].
\end{split}
\ee
where we made use of the identity of Eq.~(\ref{eq:zetaxp1}).

\bibliographystyle{JHEP}
\bibliography{hrgB}

\providecommand{\href}[2]{#2}\begingroup\raggedright\begin{thebibliography}{10}

\bibitem{Teaney:2000cw}
D.~Teaney, J.~Lauret, and E.~V. Shuryak, {\it {Flow at the SPS and RHIC as a
  quark gluon plasma signature}},  {\em Phys. Rev. Lett.} {\bf 86} (2001)
  4783--4786, [\href{http://xxx.lanl.gov/abs/nucl-th/0011058}{{\tt
  nucl-th/0011058}}].

\bibitem{Kolb:2003dz}
P.~F. Kolb and U.~W. Heinz, {\it {Hydrodynamic description of ultrarelativistic
  heavy-ion collisions}},  \href{http://xxx.lanl.gov/abs/nucl-th/0305084}{{\tt
  nucl-th/0305084}}.

\bibitem{Jacobs:2004qv}
P.~Jacobs and X.-N. Wang, {\it {Matter in extremis: Ultrarelativistic nuclear
  collisions at RHIC}},  {\em Prog.Part.Nucl.Phys.} {\bf 54} (2005) 443--534,
  [\href{http://xxx.lanl.gov/abs/hep-ph/0405125}{{\tt hep-ph/0405125}}].

\bibitem{Romatschke:2009im}
P.~Romatschke, {\it {New Developments in Relativistic Viscous Hydrodynamics}},
  {\em Int.J.Mod.Phys.} {\bf E19} (2010) 1--53,
  [\href{http://xxx.lanl.gov/abs/0902.3663}{{\tt arXiv:0902.3663}}]. 49 pages,
  12 figures, lecture notes in review form.

\bibitem{Skokov:2009qp}
V.~Skokov, A.~Y. Illarionov, and V.~Toneev, {\it {Estimate of the magnetic
  field strength in heavy-ion collisions}},  {\em Int. J. Mod. Phys. A} {\bf
  24} (2009) 5925, [\href{http://xxx.lanl.gov/abs/0907.1396}{{\tt
  arXiv:0907.1396}}].

\bibitem{Bzdak:2011yy}
A.~Bzdak and V.~Skokov, {\it {Event-by-event fluctuations of magnetic and
  electric fields in heavy ion collisions}},  {\em Phys.Lett.} {\bf B710}
  (2012) 171--174, [\href{http://xxx.lanl.gov/abs/1111.1949}{{\tt
  arXiv:1111.1949}}].

\bibitem{Deng:2012pc}
W.-T. Deng and X.-G. Huang, {\it {Event-by-event generation of electromagnetic
  fields in heavy-ion collisions}},  {\em Phys.Rev.} {\bf C85} (2012) 044907,
  [\href{http://xxx.lanl.gov/abs/1201.5108}{{\tt arXiv:1201.5108}}].

\bibitem{Duncan:1992hi}
R.~C. Duncan and C.~Thompson, {\it {Formation of very strongly magnetized
  neutron stars - implications for gamma-ray bursts}},  {\em Astrophys. J.}
  {\bf 392} (1992) L9.

\bibitem{Vachaspati:1991nm}
T.~Vachaspati, {\it {Magnetic fields from cosmological phase transitions}},
  {\em Phys. Lett. B} {\bf 265} (1991) 258.

\bibitem{Kharzeev:2007jp}
D.~E. Kharzeev, L.~D. McLerran, and H.~J. Warringa, {\it {The Effects of
  topological charge change in heavy ion collisions: 'Event by event P and CP
  violation'}},  {\em Nucl.Phys.} {\bf A803} (2008) 227--253,
  [\href{http://xxx.lanl.gov/abs/0711.0950}{{\tt arXiv:0711.0950}}].

\bibitem{Fukushima:2008xe}
K.~Fukushima, D.~E. Kharzeev, and H.~J. Warringa, {\it {The Chiral Magnetic
  Effect}},  {\em Phys. Rev.} {\bf D78} (2008) 074033,
  [\href{http://xxx.lanl.gov/abs/0808.3382}{{\tt arXiv:0808.3382}}].

\bibitem{Bali:2011qj}
G.~Bali, F.~Bruckmann, G.~Endr\H{o}di, Z.~Fodor, S.~Katz, {\em et.~al.}, {\it
  {The QCD phase diagram for external magnetic fields}},  {\em JHEP} {\bf 1202}
  (2012) 044, [\href{http://xxx.lanl.gov/abs/1111.4956}{{\tt
  arXiv:1111.4956}}].

\bibitem{Bali:2011uf}
G.~Bali, F.~Bruckmann, G.~Endr\H{o}di, Z.~Fodor, S.~Katz, {\em et.~al.}, {\it
  {The finite temperature QCD transition in external magnetic fields}},  {\em
  PoS} {\bf LATTICE2011} (2011) 192,
  [\href{http://xxx.lanl.gov/abs/1111.5155}{{\tt arXiv:1111.5155}}].

\bibitem{Bernard:2006nj}
C.~Bernard, T.~Burch, C.~E. DeTar, S.~Gottlieb, L.~Levkova, {\em et.~al.}, {\it
  {QCD equation of state with 2+1 flavors of improved staggered quarks}},  {\em
  Phys.Rev.} {\bf D75} (2007) 094505,
  [\href{http://xxx.lanl.gov/abs/hep-lat/0611031}{{\tt hep-lat/0611031}}].

\bibitem{Cheng:2009zi}
M.~Cheng, S.~Ejiri, P.~Hegde, F.~Karsch, O.~Kaczmarek, {\em et.~al.}, {\it
  {Equation of State for physical quark masses}},  {\em Phys.Rev.} {\bf D81}
  (2010) 054504, [\href{http://xxx.lanl.gov/abs/0911.2215}{{\tt
  arXiv:0911.2215}}].

\bibitem{Borsanyi:2010cj}
S.~Bors\'anyi, G.~Endr\H{o}di, Z.~Fodor, A.~Jakov\'ac, S.~D. Katz, {\em
  et.~al.}, {\it {The QCD equation of state with dynamical quarks}},  {\em
  JHEP} {\bf 1011} (2010) 077, [\href{http://xxx.lanl.gov/abs/1007.2580}{{\tt
  arXiv:1007.2580}}].

\bibitem{Karsch:2003vd}
F.~Karsch, K.~Redlich, and A.~Tawfik, {\it {Hadron resonance mass spectrum and
  lattice QCD thermodynamics}},  {\em Eur. Phys. J.} {\bf C29} (2003) 549--556,
  [\href{http://xxx.lanl.gov/abs/hep-ph/0303108}{{\tt hep-ph/0303108}}].

\bibitem{Huovinen:2009yb}
P.~Huovinen and P.~Petreczky, {\it {QCD Equation of State and Hadron Resonance
  Gas}},  {\em Nucl.Phys.} {\bf A837} (2010) 26--53,
  [\href{http://xxx.lanl.gov/abs/0912.2541}{{\tt arXiv:0912.2541}}].

\bibitem{Karsch:2003zq}
F.~Karsch, K.~Redlich, and A.~Tawfik, {\it {Thermodynamics at non-zero baryon
  number density: A comparison of lattice and hadron resonance gas model
  calculations}},  {\em Phys. Lett.} {\bf B571} (2003) 67--74,
  [\href{http://xxx.lanl.gov/abs/hep-ph/0306208}{{\tt hep-ph/0306208}}].

\bibitem{Tawfik:2004sw}
A.~Tawfik, {\it {The QCD phase diagram: A comparison of lattice and hadron
  resonance gas model calculations}},  {\em Phys. Rev.} {\bf D71} (2005)
  054502, [\href{http://xxx.lanl.gov/abs/hep-ph/0412336}{{\tt
  hep-ph/0412336}}].

\bibitem{Allton:2005gk}
C.~Allton, M.~Doring, S.~Ejiri, S.~Hands, O.~Kaczmarek, {\em et.~al.}, {\it
  {Thermodynamics of two flavor QCD to sixth order in quark chemical
  potential}},  {\em Phys.Rev.} {\bf D71} (2005) 054508,
  [\href{http://xxx.lanl.gov/abs/hep-lat/0501030}{{\tt hep-lat/0501030}}].

\bibitem{Borsanyi:2012cr}
S.~Bors\'anyi, G.~Endr\H{o}di, Z.~Fodor, S.~Katz, S.~Krieg, {\em et.~al.}, {\it
  {QCD equation of state at nonzero chemical potential: continuum results with
  physical quark masses at order $mu^2$}},  {\em JHEP} {\bf 1208} (2012) 053,
  [\href{http://xxx.lanl.gov/abs/1204.6710}{{\tt arXiv:1204.6710}}].

\bibitem{Majumder:2010ik}
A.~Majumder and B.~Muller, {\it {Hadron Mass Spectrum from Lattice QCD}},  {\em
  Phys.Rev.Lett.} {\bf 105} (2010) 252002,
  [\href{http://xxx.lanl.gov/abs/1008.1747}{{\tt arXiv:1008.1747}}].

\bibitem{NoronhaHostler:2012ug}
J.~Noronha-Hostler, J.~Noronha, and C.~Greiner, {\it {Hadron Mass Spectrum and
  the Shear Viscosity to Entropy Density Ratio of Hot Hadronic Matter}},  {\em
  Phys.Rev.} {\bf C86} (2012) 024913,
  [\href{http://xxx.lanl.gov/abs/1206.5138}{{\tt arXiv:1206.5138}}].

\bibitem{kittel2004elementary}
C.~Kittel, {\em Elementary Statistical Physics}.
\newblock Dover Books on Physics Series. Dover, 2004.

\bibitem{stanley1987introduction}
H.~Stanley, {\em Introduction to Phase Transitions and Critical Phenomena}.
\newblock The International Series of Monographs on Physics Series. Oxford
  University Press, 1987.

\bibitem{Ferrer:2010wz}
E.~J. Ferrer, V.~de~la Incera, J.~P. Keith, I.~Portillo, and P.~P. Springsteen,
  {\it {Equation of State of a Dense and Magnetized Fermion System}},  {\em
  Phys.Rev.} {\bf C82} (2010) 065802,
  [\href{http://xxx.lanl.gov/abs/1009.3521}{{\tt arXiv:1009.3521}}].

\bibitem{0022-3719-15-30-017}
R.~D. Blandford and L.~Hernquist, {\it Magnetic susceptibility of a neutron
  star crust},  {\em Journal of Physics C: Solid State Physics} {\bf 15}
  (1982), no.~30 6233.

\bibitem{Bali:2013esa}
G.~Bali, F.~Bruckmann, G.~Endr\H{o}di, F.~Gruber, and A.~Sch{\"a}fer, {\it
  {Magnetic field-induced gluonic (inverse) catalysis and pressure (an)isotropy
  in QCD}},  \href{http://xxx.lanl.gov/abs/1303.1328}{{\tt arXiv:1303.1328}}.

\bibitem{Dashen:1969ep}
R.~Dashen, S.-K. Ma, and H.~J. Bernstein, {\it {S Matrix formulation of
  statistical mechanics}},  {\em Phys. Rev.} {\bf 187} (1969) 345--370.

\bibitem{Venugopalan:1992hy}
R.~Venugopalan and M.~Prakash, {\it {Thermal properties of interacting
  hadrons}},  {\em Nucl. Phys.} {\bf A546} (1992) 718--760.

\bibitem{Beringer:1900zz}
{\bf Particle Data Group} Collaboration, J.~Beringer {\em et.~al.}, {\it
  {Review of Particle Physics (RPP)}},  {\em Phys.Rev.} {\bf D86} (2012)
  010001.

\bibitem{Ferrara:1992yc}
S.~Ferrara, M.~Porrati, and V.~L. Telegdi, {\it {g = 2 as the natural value of
  the tree level gyromagnetic ratio of elementary particles}},  {\em Phys.Rev.}
  {\bf D46} (1992) 3529--3537.

\bibitem{Goldman:1972vc}
J.~T. Goldman, W.-Y. Tsai, and A.~Yildiz, {\it {Consistency of spin-one
  theory}},  {\em Phys.Rev.} {\bf D5} (1972) 1926--1930.

\bibitem{Tsai:1972iq}
W.-Y. Tsai and A.~Yildiz, {\it {Motion of charged particles in a homogeneous
  magnetic field}},  {\em Phys.Rev.} {\bf D4} (1971) 3643--3648.

\bibitem{landau1977quantum}
L.~Landau and E.~Lifshits, {\em Quantum Mechanics: Non-Relativistic Theory}.
\newblock Course of theoretical physics (Vol 3.) (Landau, L. D, 1908-1968).
  Pergamon Press, 1977.

\bibitem{Rarita:1941mf}
W.~Rarita and J.~Schwinger, {\it {On a theory of particles with half integral
  spin}},  {\em Phys.Rev.} {\bf 60} (1941) 61.

\bibitem{Johnson:1960vt}
K.~Johnson and E.~Sudarshan, {\it {Inconsistency of the local field theory of
  charged spin 3/2 particles}},  {\em Annals Phys.} {\bf 13} (1961) 126--145.

\bibitem{Velo:1969bt}
G.~Velo and D.~Zwanziger, {\it {Propagation and quantization of
  Rarita-Schwinger waves in an external electromagnetic potential}},  {\em
  Phys.Rev.} {\bf 186} (1969) 1337--1341.

\bibitem{kapusta2006finite}
J.~Kapusta and C.~Gale, {\em Finite-Temperature Field Theory: Principles and
  Applications}.
\newblock Cambridge monographs on mechanics and applied mathematics. Cambridge
  University Press, 2006.

\bibitem{Fraga:2008qn}
E.~S. Fraga and A.~J. Mizher, {\it {Chiral transition in a strong magnetic
  background}},  {\em Phys.Rev.} {\bf D78} (2008) 025016,
  [\href{http://xxx.lanl.gov/abs/0804.1452}{{\tt arXiv:0804.1452}}].

\bibitem{Schwinger:1951nm}
J.~S. Schwinger, {\it {On gauge invariance and vacuum polarization}},  {\em
  Phys.Rev.} {\bf 82} (1951) 664--679.

\bibitem{Elmfors:1993bm}
P.~Elmfors, D.~Persson, and B.-S. Skagerstam, {\it {Real time thermal
  propagators and the QED effective action for an external magnetic field}},
  {\em Astropart.Phys.} {\bf 2} (1994) 299--326,
  [\href{http://xxx.lanl.gov/abs/hep-ph/9312226}{{\tt hep-ph/9312226}}].

\bibitem{Dunne:2004nc}
G.~V. Dunne, {\it {Heisenberg-Euler effective Lagrangians: Basics and
  extensions}},  \href{http://xxx.lanl.gov/abs/hep-th/0406216}{{\tt
  hep-th/0406216}}.

\bibitem{Andersen:2011ip}
J.~O. Andersen and R.~Khan, {\it {Chiral transition in a magnetic field and at
  finite baryon density}},  {\em Phys.Rev.} {\bf D85} (2012) 065026,
  [\href{http://xxx.lanl.gov/abs/1105.1290}{{\tt arXiv:1105.1290}}].

\bibitem{Chakrabarty:1996te}
S.~Chakrabarty, {\it {Quark matter in strong magnetic field}},  {\em Phys.Rev.}
  {\bf D54} (1996) 1306--1316,
  [\href{http://xxx.lanl.gov/abs/hep-ph/9603406}{{\tt hep-ph/9603406}}].

\bibitem{Ebert:1999ht}
D.~Ebert, K.~Klimenko, M.~Vdovichenko, and A.~Vshivtsev, {\it {Magnetic
  oscillations in dense cold quark matter with four fermion interactions}},
  {\em Phys.Rev.} {\bf D61} (2000) 025005,
  [\href{http://xxx.lanl.gov/abs/hep-ph/9905253}{{\tt hep-ph/9905253}}].

\bibitem{Agasian:2008tb}
N.~Agasian and S.~Fedorov, {\it {Quark-hadron phase transition in a magnetic
  field}},  {\em Phys.Lett.} {\bf B663} (2008) 445--449,
  [\href{http://xxx.lanl.gov/abs/0803.3156}{{\tt arXiv:0803.3156}}].

\bibitem{Menezes:2008qt}
D.~Menezes, M.~Benghi~Pinto, S.~Avancini, A.~Perez~Martinez, and
  C.~Providencia, {\it {Quark matter under strong magnetic fields in the
  Nambu-Jona-Lasinio Model}},  {\em Phys.Rev.} {\bf C79} (2009) 035807,
  [\href{http://xxx.lanl.gov/abs/0811.3361}{{\tt arXiv:0811.3361}}].

\bibitem{Boomsma:2009yk}
J.~K. Boomsma and D.~Boer, {\it {The Influence of strong magnetic fields and
  instantons on the phase structure of the two-flavor NJL model}},  {\em
  Phys.Rev.} {\bf D81} (2010) 074005,
  [\href{http://xxx.lanl.gov/abs/0911.2164}{{\tt arXiv:0911.2164}}].

\bibitem{Fraga:2012fs}
E.~S. Fraga and L.~F. Palhares, {\it {Deconfinement in the presence of a strong
  magnetic background: an exercise within the MIT bag model}},  {\em Phys.Rev.}
  {\bf D86} (2012) 016008, [\href{http://xxx.lanl.gov/abs/1201.5881}{{\tt
  arXiv:1201.5881}}].

\bibitem{Blaizot:2012sd}
J.-P. Blaizot, E.~S. Fraga, and L.~F. Palhares, {\it {Effect of quark masses on
  the QCD presssure in a strong magnetic background}},
  \href{http://xxx.lanl.gov/abs/1211.6412}{{\tt arXiv:1211.6412}}.

\bibitem{itzykson2006quantum}
C.~Itzykson and J.~Zuber, {\em Quantum Field Theory}.
\newblock Dover Books on Physics. Dover Publications, 2006.

\bibitem{Gusynin:1995nb}
V.~P. Gusynin, V.~A. Miransky, and I.~A. Shovkovy, {\it {Dimensional reduction
  and catalysis of dynamical symmetry breaking by a magnetic field}},  {\em
  Nucl. Phys. B} {\bf 462} (1996) 249,
  [\href{http://xxx.lanl.gov/abs/hep-ph/9509320}{{\tt hep-ph/9509320}}].

\bibitem{Bali:2012zg}
G.~Bali, F.~Bruckmann, G.~Endr\H{o}di, Z.~Fodor, S.~Katz, {\em et.~al.}, {\it
  {QCD quark condensate in external magnetic fields}},  {\em Phys.Rev.} {\bf
  D86} (2012) 071502, [\href{http://xxx.lanl.gov/abs/1206.4205}{{\tt
  arXiv:1206.4205}}].

\bibitem{Bali:2012jv}
G.~Bali, F.~Bruckmann, M.~Constantinou, M.~Costa, G.~Endr\H{o}di, {\em
  et.~al.}, {\it {Magnetic susceptibility of QCD at zero and at finite
  temperature from the lattice}},  {\em Phys.Rev.} {\bf D86} (2012) 094512,
  [\href{http://xxx.lanl.gov/abs/1209.6015}{{\tt arXiv:1209.6015}}].

\bibitem{Agasian:2001hv}
N.~O. Agasian, {\it {Chiral thermodynamics in a magnetic field}},  {\em Phys.
  Atom. Nucl.} {\bf 64} (2001) 554--560,
  [\href{http://xxx.lanl.gov/abs/hep-ph/0112341}{{\tt hep-ph/0112341}}].

\bibitem{Cohen:2007bt}
T.~D. Cohen, D.~A. McGady, and E.~S. Werbos, {\it {The Chiral condensate in a
  constant electromagnetic field}},  {\em Phys.Rev.} {\bf C76} (2007) 055201,
  [\href{http://xxx.lanl.gov/abs/0706.3208}{{\tt arXiv:0706.3208}}].

\bibitem{Chernodub:2010qx}
M.~Chernodub, {\it {Superconductivity of QCD vacuum in strong magnetic field}},
   {\em Phys.Rev.} {\bf D82} (2010) 085011,
  [\href{http://xxx.lanl.gov/abs/1008.1055}{{\tt arXiv:1008.1055}}].

\bibitem{Fraga:2012ev}
E.~S. Fraga, J.~Noronha, and L.~F. Palhares, {\it {Large Nc Deconfinement
  Transition in the Presence of a Magnetic Field}},
  \href{http://xxx.lanl.gov/abs/1207.7094}{{\tt arXiv:1207.7094}}.

\bibitem{peskin1995introduction}
M.~Peskin and D.~Schroeder, {\em An Introduction To Quantum Field Theory}.
\newblock The Advanced Book Program. Basic Books, 1995.

\bibitem{dlmf}
 Digital Library of Mathematical Functions, release date 2012-03-23. National
  Institute of Standards and Technology, http://dlmf.nist.gov/.

\end{thebibliography}\endgroup

\end{document}